\documentclass[12pt]{article}

\usepackage[letterpaper,hmargin=1in,vmargin=1in]{geometry}

\usepackage{graphicx,epstopdf,amsmath,amsfonts}

\def\be{\begin{equation}}
\def\ee{\end{equation}}
\def\ba{\begin{eqnarray}}
\def\ea{\end{eqnarray}}
\def\ge{\mathrel{\raise.3ex\hbox{$>$\kern-.75em\lower1ex\hbox{$\sim$}}}}
\def\la{\mathrel{\raise.3ex\hbox{$<$\kern-.75em\lower1ex\hbox{$\sim$}}}}

\def\simgt{\mathrel{\raise.3ex\hbox{$>$\kern-.75em\lower1ex\hbox{$\sim$}}}}
\def\simlt{\mathrel{\raise.3ex\hbox{$<$\kern-.75em\lower1ex\hbox{$\sim$}}}}

\newcommand{\bi}[1]{\bibitem{#1}}
\newcommand{\fr}[2]{\frac{#1}{#2}}

\newcommand{\nc}{\newcommand}

\nc{\gone}{\bar g_{\pi NN}^{(1)}}
\nc{\gzero}{\bar g_{\pi NN}^{(0)}}
\nc{\al}{\alpha}
\nc{\ga}{\gamma}
\nc{\de}{\delta}
\nc{\ep}{\epsilon}
\nc{\ze}{\zeta}
\nc{\et}{\eta}
\nc{\ka}{\kappa}
\nc{\rh}{\rho}
\nc{\si}{\sigma}
\nc{\ta}{\tau}
\nc{\up}{\upsilon}
\nc{\ph}{\phi}
\nc{\ch}{\chi}
\nc{\ps}{\psi}
\nc{\om}{\omega}
\nc{\Ga}{\Gamma}
\nc{\De}{\Delta}
\nc{\La}{\Lambda}
\nc{\Si}{\Sigma}
\nc{\Up}{\Upsilon}
\nc{\Ph}{\Phi}
\nc{\Ps}{\Psi}
\nc{\Om}{\Omega}
\nc{\ptl}{\partial}
\nc{\del}{\nabla}
\nc{\ov}{\overline}
\nc{\newcaption}[1]{\centerline{\parbox{15cm}{\caption{#1}}}}

\def\beq{\begin{equation}}
\def\eeq{\end{equation}}
\def\bmat{\begin{displaymath}}
\def\emat{\end{displaymath}}
\def\bear{\begin{eqnarray}}
\def\eear{\end{eqnarray}}
\def\ba{\begin{eqnarray}}
\def\ea{\end{eqnarray}}
\def\bery{\begin{array}}
\def\ery{\end{array}}
\def\bit{\begin{itemize}}
\def\eit{\end{itemize}}
\def\ben{\begin{enumerate}}
\def\een{\end{enumerate}}
\def\btab{\begin{tabular}}
\def\etab{\end{tabular}}
\def\btbl{\begin{table}}
\def\etbl{\end{table}}
\def\bfig{\begin{figure}[htb]}
\def\efig{\end{figure}}
\def\bpic{\begin{picture}}
\def\epic{\end{picture}}

\def\ga{\mathrel{\raise.3ex\hbox{$>$\kern-.75em\lower1ex\hbox{$\sim$}}}}
\def\la{\mathrel{\raise.3ex\hbox{$<$\kern-.75em\lower1ex\hbox{$\sim$}}}}
\def\gappeq{\mathrel{\rlap {\raise.5ex\hbox{$>$}}
{\lower.5ex\hbox{$\sim$}}}}
\def\lappeq{\mathrel{\rlap{\raise.5ex\hbox{$<$}}
{\lower.5ex\hbox{$\sim$}}}}

\def\gyr{{\rm \, G\kern-0.125em yr}}
\def\mev{{\rm \, Me\kern-0.125em V}}
\def\gev{{\rm \, Ge\kern-0.125em V}}
\def\tev{{\rm \, Te\kern-0.125em V}}

\begin{document}

\begin{titlepage}

\setcounter{page}{1}

\vspace*{0.2in}

\begin{center}

\hspace*{-0.6cm}\parbox{17.5cm}{\Large \bf \begin{center}
The galactic 511\,keV line from electroweak scale WIMPs\end{center}}

\vspace*{0.5cm}
\normalsize

{\bf  Maxim Pospelov$^{\,(a,b)}$ and Adam Ritz$^{\,(a)}$}

\smallskip
\medskip

$^{\,(a)}${\it Department of Physics and Astronomy, University of Victoria, \\
     Victoria, BC, V8P 1A1 Canada}

$^{\,(b)}${\it Perimeter Institute for Theoretical Physics, Waterloo,
ON, N2J 2W9, Canada}

\smallskip
\end{center}
\vskip0.2in

\centerline{\large\bf Abstract}

We consider possible mechanisms via which electroweak scale WIMPs $\ch^0$ could provide the source of the INTEGRAL/SPI
511\,keV photon flux from the galactic centre. We consider scenarios where the WIMP spectrum contains near-degeneracies,
with MeV-scale splitting, and focus on three possible production mechanisms for galactic positrons: (i) collisional excitation
of the WIMP to a nearby charged state, $\ch^0 + \ch^0 \rightarrow \ch^+ + \ch^-$, with the subsequent decay producing
positrons; (ii) capture of the WIMP by nuclei in the galactic interstellar medium, $\ch^0 + N \rightarrow e^+ + (\ch^- N)$; and (iii) the
decay of a nearby long-lived state surviving from the Big Bang, $\ch^0_2 \rightarrow \ch_1^0 + e^+ + e^-$. We find that process (i) 
requires a cross-section which is significantly larger than the unitarity bound, process (ii) is allowed by unitarity, but is impractical due to 
terrestrial bounds on the $\ch N$ cross-section,  while process (iii) is viable and we construct a simple model realization
with singlet dark matter fields interacting with the Standard Model via the Higgs sector.

\vfil
\leftline{March 2007}

\end{titlepage}

\section{Introduction}

Given what is now rather compelling evidence for the existence of dark matter on various astrophysical and
cosmological scales, finding direct and/or indirect means for discerning its nature stands as one of the
most profound problems in particle physics. In this regard, the use of local astrophysical observations as an
indirect probe is increasingly becoming one of our most powerful tools, and there are a number of galactic observations
which can provide us with useful guidance and constraints on the particle physics of dark matter. 

As a prominent example, the 511~keV line from the central region of our galaxy, now well-measured 
using the SPI spectrometer on the INTEGRAL satellite \cite{SPI1,SPI2}, represents something of a
challenge for theoretical astrophysics. Indeed, the strength of the signal, a photon flux of 
\be 
 \Ph_{\rm exp} = (9.35 \pm 0.54)\times 10^{-4} \; {\rm ph}\,{\rm cm}^{-2}\, {\rm s}^{-1},
 \ee
 and in particular its large component from the galactic bulge, is rather unexpected given the
 known sources and production mechanisms for galactic positrons. A number of possible explanations for 
 the INTEGRAL/SPI signal using conventional Standard Model (SM) physics along with somewhat novel astrophysics have been 
 put forward (see \cite{SPI2} and references therein), of which production within Type Ia supernovae and/or  
 low-mass X-ray binaries appear the most
 plausible \cite{SPI2}, but have considerable difficulty in explaining the enhanced bulge component. The apparent localization of the 
 source to the inner 1--2~kpc of the galactic core has also led to numerous speculations on the possibility of 
 a  non-standard origin for the  511~keV line, or more precisely for the positrons that fuel the annihilation. Since 
 the density of  dark matter is expected to be significantly enhanced in the 
central region \cite{NFW}, it is indeed tempting to link this line to dark matter particles, and `explain' the signal via 
self-annihilation,  decay, or interaction with baryonic matter. The dark-matter-based scenarios that
have been put forward include:  the annihilation of MeV-scale weakly-interacting massive particles (WIMPs) \cite{MeVann};
the decays of super-weakly interacting MeV-scale dark matter particles \cite{pp,HW}; along with more intricate mechanisms such as
the annihilation of electrons inside clumps of antimatter surviving from the Big Bang \cite{Oaknin:2004mn};
and a superconducting network of cosmic strings producing positrons in the magnetic field \cite{Ferrer:2005xv}.
Further discussion of these ideas and other models that mostly  follow the aforementioned possibilities can be 
found in \cite{Arlington}. 

Many of the dark matter models suggested as a source of the 511~keV line require some 
intricate model building, as discussed in some detail in \cite{BF}. An ${\cal O}$(MeV) scale dark matter particle is well 
outside the expected WIMP mass range \cite{LW}. Thus, such models typically 
require the addition of a new force mediating the interaction between dark matter and 
the Standard Model sector, with the mass of the mediator significantly lighter than the electroweak scale. 
In this paper we will analyze several options for producing galactic positrons from dark matter
particles with masses in the conventional range for WIMPs at or around the electroweak scale, {\em e.g.} from several tens of GeV to 
several TeV. Models of WIMPs in this mass range are considerably easier to build than 
${\cal O}$(MeV) scale dark matter, and thus at the outset appear more natural.

At first sight, there are some significant constraints on producing $O$(MeV)-energy positrons from
${\cal O}$(TeV) scale WIMPs. Indeed, an important constraint on the energy of the injected positrons can be deduced from the 
line shape of the 511~keV signal \cite{shape}, and from considerations of 
the MeV $\gamma$-rays accompanying the decay or annihilation of dark matter 
particles \cite{Beacom1,Beacom2}. This requires that the 
spectrum of positrons be rather soft, with injection energies 
as low as ${\cal O}$(10) MeV or less.  This clearly rules out direct annihilation of TeV-scale WIMPs as a possible source of 
the galactic 511 keV line, as the annihilation products would typically be very energetic,
carrying away a significant fraction of the WIMP mass. However, this impasse is clearly circumvented if there exist some
additional heavier neutral or charged states, $\ch_2^0$ or $\ch^+_2$, nearly degenerate with the dark matter
state, that we will generically denote $\ch_1^0$. In particular, given a suitable  mass splitting, positrons can 
 originate in transitions between these states,
\begin{eqnarray}
\label{charged}
 &&\chi^+_2\to \chi^0_1+ e^++\cdots\\
 &&\chi^0_2\to \chi^0_1 + e^+ +e^- +\cdots
\label{neutral}
\end{eqnarray}
If the mass difference is in the MeV-range, 
$m_e< m_{\chi^+_2}-m_{\chi^0_1}\la 5$MeV or $2m_e< m_{\chi^0_2}-m_{\chi^0_1}\la 10$MeV, the positrons 
emerging from the decay of $\chi^+_2$ and/or $\chi^0_2$ would be in the required 
energy range to fit the line shape of the 511~keV signal, and 
comply with the bounds on the $\gamma$-spectrum in the MeV-range. 

While this possibility of having a near-degeneracy in the dark sector may seem somewhat {\em ad hoc} in its application
to the 511~keV line, such a possibility is actually independently motivated by a number of theoretical issues pertaining
to WIMP physics.
Perhaps most prominently, the near degeneracy of neutral and charged states, with $\Delta m\ll m$, 
allows for more efficient depletion of the neutral state abundance (coannihilation) at freeze-out \cite{GS}. This
requirement is particularly apparent in supersymmetric models, where much of the viable parameter space
requires enhanced depletion mechanisms to avoid over-closure of the Universe; a well-known example
being coannihilation of the neutralino LSP with a nearly degenerate stau. The decay of one dark matter state 
into another has also been proposed as a possible solution to the over-clumping of cold dark matter 
on sub-galactic distance scales \cite{caltech}, while it has been  suggested that the substructure of WIMPs might be 
responsible for the DAMA signal \cite{DAMA} if the cross section is dominated by an inelastic collision \cite{SW}. 
Finally, the existence of metastable charged dark matter companions can lead to interesting modifications of the primordial elemental 
abundances via the catalysis of the Big Bang Nucleosynthesis \cite{CBBN1,CBBN2}. However, while these ideas provide ample
motivation for considering near-degeneracies, one has to bear in mind that in many of these proposals 
$\Delta m$ is not required to be in the MeV range, as {\em e.g.} an $O$(GeV)
splitting would already enable the co-annihilation of neutralinos and scalar leptons, and thus 
for all models that we consider in this paper the fine-tuning of $\De m$ is an additional {\em ad hoc} requirement.  

A second requirement is that the transitions between $\chi^+_2$, $\chi^0_2$  and $\chi^0_1$ 
should happen sufficiently often to provide the positron flux needed to account for the 
strength of the 511 KeV signal. This requirement translates into different types of 
constraints for models where $\chi^+_2$ and/or $\chi^0_2$ survive from the Big Bang, and for models 
where these states are produced in collisions between dark matter particles and/or dark matter 
particles and nuclei. For the case of collisional excitation, 
the cross section for producing the excited states should be rather 
large to compensate for the low number density of dark matter,
typically in the range of $ 10^{-3}/$cm$^3$. In models where the
excited states of dark matter survive from the Big Bang, the positron flux imposes 
both upper and lower limits on the lifetimes of these excited states. 

The layout of this paper is as follows.
In the next three sections (Sections 2-4) we analyze the feasibility of three generic scenarios:
\begin{itemize}
\item[(i)] Collisional excitation of WIMPS, with the subsequent decay producing positrons.
\item[(ii)]  Emission of positrons in WIMP-nucleus recombination.
\item[(iii)]  Decay of metastable WIMPs surviving from the Big Bang to their ground states.
\end{itemize}
Our conclusions are that for case (i) the required cross section for the $\ch_1^0$-$\ch_1^0$ collision has to be very large, 
well in excess of the unitarity bound for any partial wave $l\sim {\cal O}(1)$, which essentially provides a model-independent argument ruling out
the possibility of producing sufficient galactic positrons from collisional excitations of WIMPs. It appears that the only way around this conclusion
is if the cross-section is saturated by large $l$'s, so that it assumes the Rutherford form (which would require neutral 
excited states), but this seems very unlikely given the fact that the collision is necessarily highly inelastic for generic WIMPs.
For case (ii), although WIMP-nucleus recombination cannot be ruled 
out as a source of $e^+$ purely on the grounds of unitarity, the required size of the cross section 
tends to violate the constraints on the anomalous abundance of heavy isotopes. The final possibility (iii),  
the decay of excited metastable WIMP states surviving from the Big Bang, must occur with a  
lifetime interval of  between $\sim 10^9$ and $10^{13}$ years. Thus, this is the only clearly viable option 
analyzed in this work, and we identify a broad class of  dark matter models that comply with this requirement. 
We reach our conclusions in Section~5.

{\bf Note added:} While this paper was being finalized we became aware of 
recent work by Finkbeiner and Weiner \cite{FW} which 
also considers the possibility (i) of producing positrons 
for the 511~keV line via collisional excitation of WIMPs, and thus
overlaps with Section~2. Their conclusions differ on the viability of 
this mechanism, but their calculation of the excitation cross section 
in a specific WIMP model extends the first-order Born formula beyond its range of validity, and is seemingly 
in contradiction with the unitarity bound discussed in Section~2. As noted above, since their model 
involves neutral excited states, it could
in principle be viable given a cross-section saturated by large-$l$ peripheral scattering, but this appears a
challenging model building problem given the need for a highly inelastic collision.

\section{Collisional excitation in WIMP-WIMP scattering}

\begin{figure}
\centerline{\includegraphics[bb=0 400 600 740, clip=true, width=10cm]{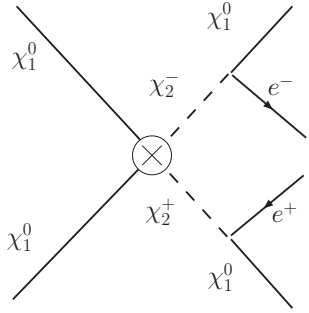}}
\vspace*{-0.3in}
 \caption{\footnotesize Collisional excitation producing two charged states $\ch_2^{\pm}$ and their subsequent decay back to $\ch_1^0$ producing
 positrons. }
\label{f1} 
\end{figure}

A TeV-mass particle moving with velocity $v\sim O(10^{-3})$ has a kinetic energy comparable to the 
electron mass. Therefore it can kinematically excite another closely degenerate state, as shown in 
Figure \ref{f1}. This process has a kinematic cutoff, i.e. in the center-of-mass frame,
\be
  v  \geq \sqrt{\frac{2m_e}{m_0}},
  \label{kin}
  \ee
 where $m_0$ is the WIMP mass, such that the excited state has mass $m_+> m_0 + m_e$.
Thus only the more energetic part of the dark matter velocity spectrum can participate in the reaction. 

To proceed, we utilize the maximal inelastic cross section allowed by unitarity, which for
the $l^{th}$ partial wave takes the form \cite{LL},
\be
\sigma_{\rm max} =\fr{\pi(2l+1)}{k_{\rm cm}^2}= \fr{\pi(2l+1)}{m_0^2v^2},
\label{unitarity}
\ee
where $k_{\rm cm}=\fr{m_0}{2}(2v)$ is the center-of-mass momentum. Obviously, outside the 
kinematic range (\ref{kin}) the cross section is zero. It is instructive to make an estimate of 
the size of the collisional rate given by this cross section for $v\sim O(10^{-3})$, and $m_0 \sim 1$TeV:
$\langle \sigma v\rangle \la 10^{-30}$cm$^2$. 

In this paper, we will simply assume a 
Maxwellian velocity distribution, noting that the actual physical distribution should implement a
sharp cutoff above the escape velocity,
\be
 \langle X(v) \rangle = \frac{4}{\sqrt{\pi}v_0^3}\int_0^\infty dv\,X(v)\, v^2 \exp(-v^2/v_0^2),
 \label{distrib}
 \ee
 such that  $\langle v^2 \rangle = 3 v_0^2/2 \equiv v_{\rm rms}^2$.
This distribution actually {\em overestimates} the more energetic fraction of dark matter, 
making the realistic velocity average of the reaction rate even smaller than the following,
\be
\langle \sigma_{\rm max}v_{\rm rel}  \rangle = \fr{2\pi(2l+1)}{m_0^2}\left\langle \fr1v \right\rangle 
= \fr{4\sqrt{\pi}(2l+1)}{m_0^2v_0}\exp\left(-\fr{2m_e}{m_0v_0^2}\right),
\label{rate}
\ee
where we have imposed the kinematic cutoff (\ref{kin}).

We are now in a position to ask if the rate (\ref{rate}) is sufficient to produce enough $\chi^+$ states for the positrons 
emerging from the subsequent decay back to $\chi_0$ to saturate the experimentally observed flux.  The required 
cross section $\langle \sigma v \rangle$ was obtained in \cite{BJ} using several density profiles,
\be
\langle \sigma v \rangle_{\rm exp} \simeq 10^{-28}{\rm~cm^2}\times \left(\fr{m_0}{1{\rm TeV}}\right)^2.
\label{needed}
\ee
In the analysis of \cite{BJ}, this value of $\langle \sigma v\rangle$ was obtained using the 
NFW density profiles \cite{NFW}, but other steeper models would produce an enhancement factor of a few at most. 
Moreover, the velocity profiles usually get softer \cite{BJ} as one moves to the central 
region. We will assume the rms velocity remains constant and use $v_{\rm rms} \simeq 220$km/s, which again 
can only result  in overestimating the flux of positrons. 

It is convenient to define the ratio $R(m_0,l)$, given by normalizing the maximal rate (\ref{rate}) 
by that required for the experimentally observed flux (\ref{needed}), 
\be
R(m_0,l)\equiv  \fr{\langle \sigma_{\rm max}v_{\rm rel} \rangle }{\langle \sigma v \rangle_{\rm exp}}= 
0.04\times \fr{4\sqrt{\pi}(2l+1)}{m_{100}^4v_0}\exp\left(-\fr{2m_e}{m_0v_0^2}\right),
\label{Rml}
\ee
where $m_{100}$ is the normalized WIMP mass $m_0/$(100 GeV). We will now argue that $R(m_0,l)$ remains smaller than 
one for any value of the WIMP mass, and any realistic partial wave $l$, thus making collisional WIMP-WIMP excitation impractical as
a source of galactic positrons sufficient for the 511~keV line. This is straightorwardly achieved by maximizing the ratio 
$R(m_0,l)$ over the mass  $m_0$ of dark matter particles. 
Indeed, the $m_0$-dependence of (\ref{Rml}) has a sharp maximum and the  
optimal mass value is around 700 GeV,
\be
m_0^{\rm optimal} = \frac{m_e}{2v_0^2} \simeq 710~{\rm GeV}\times \left(\fr{220~{\rm km/s}}{v_{\rm rms}}\right)^2.
\ee
The maximal ratio is still much smaller than 1 for any realistic value of the 
partial wave $l$:
\be
R(m_0^{\rm optimal},l) = 3.4\times 
10^{-3}\times (2l+1)\left(\fr{v_{\rm rms}}{220~{\rm km/s}}\right)^7.
\label{notenough}
\ee
This conclusion negates the hope of explaining the 511~keV line through collisional excitations in 
WIMP-WIMP scattering. In fact, the result (\ref{notenough}) implies that collisional excitation 
of WIMPs would be a subdominant source of positrons even in comparison with various conventional 
sources of $e^+$ that could account for up to a few percent of the INTEGRAL/SPI signal.  

In considering possible routes around this `no-go' theorem, its worth recalling  that the total cross section
can sometimes be much larger than the individual partial-wave cross sections for  low $l$. There is no significant
enhancement for the charged excited states considered above, but there can be in the case of neutral states. The 
most famous example is, of course, the Rutherford formula where large $l$'s, or equivalently 
large impact parameters, dominate the sum over partial waves and provide an additional enhancement 
at small scattering angles, or more precisely at small momentum transfer $q$. The maximal enhancement can be 
obtained for a Yukawa-type potential, $\exp(-r/\lambda)/r$, where $\lambda$ is the Compton wavelength of the 
force carrier. The scaling of the differential cross section with $q$ 
takes the following form: $d\sigma\sim dq^2/(q^2+\lambda^{-2})^2$, and for an arbitrarily large range $\lambda$ 
the total {\em elastic} cross section would diverge. Accounting for the Coulomb factor, $(\alpha/v)^2$, in this
case indicates that the cross-section may grow as large as $\si \sim m_0^2/q_{\rm min}^4$ in the Rutherford regime
which may give a significant enhancement if the collision is highly peripheral, naively enough to provide the requisite 
flux according to (\ref{needed}). 
However, in the present case, 
the kinetic energy in the center-of-mass frame for a 1~TeV WIMP is of the same order as the energy transfer
required to produce the excited state, and thus the collision is highly inelastic.  For inelastic processes with finite energy transfer $\Delta E$,
the scattering at zero angle has an associated minimal momentum transfer $q_{\rm min} = \Delta E/v$ \cite{LL}, and significantly
since the collision is highly inelastic, its hard to see how the cross-section could be dominated by highly peripheral collisions. 
We would therefore expect that this enhancement would not be possible, and that the inelastic nature of the collision would
require $l\sim O(1)$, leading to a cross-section no larger than $1/q_{\rm min}^2$ suggesting a maximal rate $\langle \sigma v \rangle$ 
in the ballpark of $10^{-30}$cm$^2$ as obtained above, given $q_{\rm min} \sim$ 1~GeV. Nonetheless, we leave this possibility of 
large-$l$ enhancement open as an intriguing model-building challenge.\footnote{We thank Neal Weiner for interesting correspondence on the possibility of enhancements of this form in regard to the model discussed in \cite{FW}.}

\section{Positron emission in WIMP-nucleus recombination}

\begin{figure}
\centerline{\includegraphics[bb=0 400 600 740, clip=true, width=12cm]{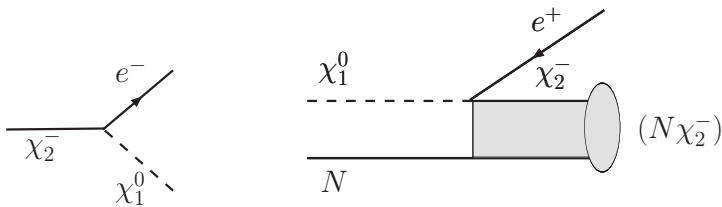}}
\vspace*{-1.3in}
 \caption{\footnotesize One the left, the rapid decay of the excited charged state $\ch_2^-$, and on the right, the capture process of $\ch_1^0$ forming the
 bound state $(N\ch_2^-)$ and producing a positron. }
\label{f2} 
\end{figure}

Collisional excitations of WIMPs with nuclei suffer from the same constraints as in the previous section,
as the abundances of very heavy nuclei with masses in excess of 100 GeV are even lower than for dark matter
particles. However, there is  a new way of emitting positrons in the process of `recombination' of {\em light} 
nuclei such as C, N, and O with WIMPs. Recombination may happen if there are charged states 
$\chi^+_2$ and $\chi^-_2$ near in mass to the WIMP state $\chi_1^0$. Naively, a transition between charged and neutral 
states of $\chi$ could occur due to weak currents, which would suppress all relevant cross sections to a level 
inconsequential for the 511~keV line. If, however, the charged states represent bosons and the neutral 
states fermions, or vice versa, the processes shown in Figure~\ref{f2} become possible. In particular, the charged states
are rapidly depleted in the early universe via processes such as $\chi^\pm_2 \to \chi^0_1+ e^\pm$, while in the current
epoch positrons may be produced by `recombination' of the form,
\be
\chi^0_1+ N \to (N \chi^-_2) +e^+,
\label{X0capture}
\ee
where $N$ denotes a generic nucleus, and $(N \chi^-_2)$ is a stable bound state of 
nuclei with the negatively charged particles. 

A straightforward quantum mechanical calcuation determines the allowed range of the mass splitting for such processes 
to become energetically possible for {\em all} incoming particle energies down to $E=0$:
\begin{eqnarray}
{\rm recombination~with~ ^{12}C}:~~~~~ 0.511~{\rm MeV} < m_{\chi^-_2} - m_{\chi^0_1}<2.30~{\rm MeV}\phantom{,}\\
{\rm recombination~with~ ^{14}N}:~~~~~ 0.511~{\rm MeV} < m_{\chi^-_2} - m_{\chi^0_1} <2.96~{\rm MeV}\phantom{,}\\
{\rm recombination~with~ ^{16}O}:~~~~~ 0.511~{\rm MeV} < m_{\chi^-_2} - m_{\chi^0_1} <3.50~{\rm MeV}.
\end{eqnarray}
In this calculation, we used a Gaussian charge distribution inside the nucleus taking 2.47, 2.55 and 2.70 
fm for the rms charge radii of C, N, and O, and implemented the limit  $m_{\chi}\gg m_{N}$. Recombination with lighter elements 
such as Li, Be, and B is also possible but perhaps less interesting in this context as these elements are not produced in stars and thus 
found only in small abundances. More importantly, recombination with H and $^4$He is not possible when accompanied by 
the emission of a positron, as the binding energy in the latter case is just 350~keV.

We can now estimate the size of the $\chi-N$ cross section needed to produce the required positron flux. If the 
emission of positrons in WIMP annihilation/excitation at any given distance $r$ from 
the center of the galaxy is proportional to $n_{DM}^2(r)$, the rate of recombination with nuclei 
scales according to $n_{DM}(r)\times n_N(r)$, where $n_N$ and $n_{DM}$ are the nuclear and WIMP number densities. 
It is expected that $n_N(r)$ is slightly enhanced in the bulge region 
relative to  known solar  abundances \cite{abundances}, and to get an estimate for the required rate we adopt
$10^{-3}$cm$^{-3}$ as a constant number density for 
C, N or O in the central region, noting that a more sophisticated model for the 
abundance could be implemented. The resulting 511~keV photon flux, normalized to that 
observed can then be expressed as \cite{pp}\footnote{This formula correctly accounts for a factor of  $1/4$, the relative probability of 
producing para-positronium,  mistakenly omitted in Ref. \cite{pp}.}:
\be
 \frac{\Ph(m_\chi,\langle \sigma_{\rm rec} v\rangle)}{\Ph_{\rm exp}} \sim  
\frac{\langle \sigma_{\rm rec} v\rangle}{10^{-28}\,{\rm cm^2}} \times \left(\frac{100\,{\rm GeV}}{m_2}\right), 
   \ee
 utilizing the density profiles of Ref.~\cite{Julio}. 
Tuning this ratio to one, one derives the  required rate for $\chi-N$ recombination to be on the order of 
$10^{-28}\,{\rm cm^2}$ for 100GeV WIMPs. This number 
   is not dramatically different from (\ref{needed}), because the abundances of C, N, O and 
   the WIMPs are of the same order. 
   However, it is important to recognize that the unitarity limit for the recombination cross section is 
   now defined by the momentum of the nucleus, which is at least one order of magnitude smaller than 
   the WIMP momentum. Consequently, we determine the maximal allowed cross section to be 
   \be
   \langle \sigma_{\rm max} v \rangle \sim 10^{-28}~{\rm cm^2},
   \ee
 which is of the same order as the required rate. Therefore, we conclude that the scenario of positrons from WIMP-nucleus 
recombination  cannot be ruled out purely on the grounds of unitarity.

However, while this may sound more promising, it appears that there are a number of terrestrial constraints
which render this scenario impractical. In particular, constraints on the abundance of heavy isotopes 
place rather stringent constraints in this case. The strongest of these, on (He$\ch_2^-$) from the absence of
heavy isotopes of hydrogen \cite{smith82}, is not relevant here since as noted above this binding does not occur
for $m_{\chi^-_2} - m_{\chi^0_1}>m_e $. However, two problematic
examples are the relative abundance constraints of $f_{\rm exp} \la 10^{-20}$ on heavy isotopes of C \cite{hemmick90}, from production of
(N$\ch_2^-)$ in the atomosphere, and $f_{\rm exp} \la  10^{-14}$ on heavy isotopes of B \cite{hemmick90}, from production of (C$\ch_2^-)$ which
allows for a significant enrichment of the anomalous B sample due to the high relative terrestrial abundance of C. 
Given that $\langle \si_{\rm rec} v \rangle < f_{\rm exp} /( n_{\rm DM} \ta_{\rm exp})$, where $\ta_{\rm exp}$ is the exposure time, the 
bound on heavy isotopes of B is sufficient  to enforce,
\be
 {\rm heavy~isotope~abundance}: \;\;\;\;\left. \langle \si_{\rm rec} v \rangle\right|_{\rm max} <  10^{-37} \, {\rm cm}^2 \left(\frac{10^9\,{\rm yr}}{\ta_{\rm exp}}\right), 
 \ee
 so that even with some imprecision in 
the exposure time, the allowed cross-section is still too small by many orders of magnitude. Note that the limits on exotic isotopes
of heavy elements, such as Fe or Au, are somewhat weaker \cite{mohapatra} but would impose a more robust constraint if e.g. binding to
lighter nuclei were disfavored, assuming the exotic nuclei, e.g. (Hg$\ch_2^-$) for Au, were able to propagate to the samples tested.

Beyond constraints on isotopic abundances, it is interesting to consider if direct terrestrial dark matter searches are sensitive
to recombination processes of this type. At first sight, it is unclear if the recoil during the capture process would be observed, e.g. in CDMS, 
due to the additional MeV-scale energy release associated with the subsequent positron annihilation. However, it is clear that if such events are
observable, the ensuing constraint on the cross-section could be even stronger than the isotopic abundance bounds discussed above. Following
a more speculative vein, one could conceive of a scenario where the binding to (I$\ch_2^{-}$), as used in DAMA \cite{DAMA}, were energetically allowed
thus enhancing the cross-section with heavy elements such as I, but 
that binding to lighter elements such as (Ge$\ch_2^-)$, as used in CDMS \cite{cdms}, was not possible. The window in mass splitting for this scenario
can be estimated as above to be 15~MeV $< \De m <$ 20~MeV, and one would then expect a signal in experiments using e.g. Xe detectors.

To complete the rather discouraging picture of the application of WIMP-nucleus recombination to the 511~keV line, it turns out that simple model
estimates of the cross-section do not in any case allow one to approach the needed rate. 
We estimate the $\chi-N$ cross section by parametrizing the strength of 
  the $\chi^0_1-\chi^-_2-e^+$ vertex by a coupling $g$. The dependence of the recombination rate on the 
parameters of this model is easily obtained, 
  \be
  \langle \sigma_{\rm rec} v \rangle \sim g^2 a_B^3 \fr{E^2_{e^+}}{ m_\chi},
  \label{modelrate}
  \ee
where $a_B$ is the typical distance between 
$\chi$ and $N$ within the bound state, $a_B\sim {\cal O}$(fm), and $E_{e^+}$ is the positron energy. 
Although the presence of $g$ as a free parameter  leaves open the 
hope of tuning this cross section to a large value, in reality any realistic choice for $g$ 
would leave the rate (\ref{modelrate}) several orders of magnitude below the required level.

\section{Delayed decays of excited WIMP states}

We turn now to the third mechanism, namely the late decay of excited WIMPs to the ground state. In this case, as we will make more
precise below, it is clear that one can produce a sufficient 511~keV flux even with a relatively low abundance as follows e.g. by rescaling
the results of \cite{pp} to the WIMP mass range.  Before turning to this discussion, we note that it is most natural to restrict our 
attention to neutral metastable states $\ch_2^0$, as the alternative option of a charged state $\ch_2^{\pm}$ is subject to stringent 
constraints. In particular, the presence of a long-lived $\ch_2^-$ state 
would alter the abundance of light elements during BBN, most notably $^6$Li \cite{CBBN1,CBBN2,CBBN3}, and the typical
constraint on the abundance relative to baryons is ${\cal O}(10^{-6})$ \cite{CBBN1,CBBN3}. While such an abundance
might barely be consistent with the required flux, most of these states will bind with He forming (He$\ch_2^-)$, and thus will be
subject to the stringent terrestrial bounds of ${\cal O}(10^{-28})$ on the fractional abundance of heavy isotopes of hydrogen \cite{smith82},
that would not allow for the required flux.

We will therefore focus on the decay $\ch_2^0 \rightarrow \ch_1^0 + e^+ +e^-$, and the
need to produce positrons with energies no larger than a few MeV implies  a mass splitting 
$\De m = m_2 - m_1 \sim {\cal O}({\rm MeV})$. 
It is straightforward to relate the required $e^+e^-$ decay width to the lifetime of the 
excited state. The line-of-sight integral of the dark matter density profile leads to the following relation for the flux  \cite{pp},
   \be
    \frac{\Ph(m_2,\Ga_{e^+e^-})}{\Ph_{\rm exp}} \approx 
(1.3 \times 10^{13}\,{\rm yr} \times \Ga_{e^+e^-}) \left(\frac{100\,{\rm GeV}}{m_2}\right), \label{flux}
   \ee
and since the total width $\Gamma_{tot} \geq \Ga_{e^+e^-}$, this immediately fixes 
the allowed window on the lifetime of $\chi^0_2$,
\be
{\rm few\times 10^9~yr}\la \tau_{\chi^0_2} \la 10^{13}~{\rm yr}.
\ee

As an illustration of the significance of this constraint, we will consider first a possible model of this type within the 
MSSM, involving a decay between the next-to-lightest and
lightest neutralino states with the mass splitting tuned as above. The next-to-lightest state then becomes metastable.
However, it is clear that a simple one-loop ({\em e.g.} chargino-stop) 
diagram contributes to the transitional dipole moment between the two neutralino states, and in the absence of 
additional mass hierarchies within the MSSM  the heavier state still has a relatively short lifetime:
\be
\Gamma_{tot} \sim \left(\fr{\alpha}{4\pi}\right)^2\fr{(\Delta m)^3}{8\pi m^2} \ga 1~{\rm sec}^{-1},
\ee
asssuming typical TeV-scale values for the neutralino masses. It is clear that the short  lifetime of $\chi^0_2$
in this case is a serious challenge for any models of non-zero spin WIMPs that interact with the charged sector
of the SM or the MSSM. Spin-zero WIMPs have a certain advantage in this sense as single photon 
transitions are forbidden. 

\begin{figure}
\centerline{\includegraphics[bb=0 400 600 740, clip=true, width=10cm]{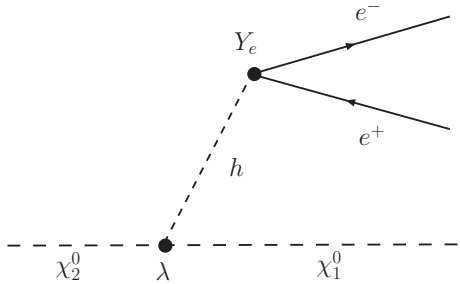}}
\vspace*{-0.3in}
 \caption{\footnotesize Higgs-mediated decay of a long-lived metastable state $\ch^0_2$ to the WIMP $\ch^0_1$ producing positrons. }
\label{f3} 
\end{figure}

We will now construct a simple model that can realize the required 511 KeV flux, assuming the $\ch^0$ states interact with the
Standard Model at tree-level only through the Higgs sector.\footnote{Assuming direct coupling to the Standard Model sector, the
only alternative mediation mechanism would be via  an axial-vector coupling to the $Z$ (a vector coupling is excluded by direct-search 
constraints on the elastic cross-section). 
In this case the $\gamma$-background would again be suppressed. However, a much smaller coupling 
would be required due to the absence of the  $m_e/v$ suppression factor noted above for Higgs mediation.}
  This model follows rather closely the minimal singlet-scalar WIMP model suggested in \cite{JM,bpv}, 
and has the minimum number of two singlet fields required for WIMP substructure. 
The parameters of the model are chosen to fit (\ref{flux}), which may appear artificial, 
but our aim here is not to construct a completely `natural' dark matter model but simply
to illustrate that this production process for the photon flux is indeed viable 
with a `proof of principle'. 

The relevant decay is shown in Fig.~\ref{f3}. If $\ch_{(1,2)}$ are scalar states, which interact with the Higgs as follows,
\be
 {\cal L}_{\rm int}^{\rm scalar} = - \lambda v h \ch^0_1 \ch^0_2,
\ee
the decay rate, assuming $\De m/m_1 \ll 1$ while still producing relativistic $e^+e^-$ pairs, is given by 
\be
 \Ga^{\rm scalar}_{e^+e^-} = \frac{m_e^2 \lambda^2}{384\pi^3 m_2} \left( \frac{\De m}{m_h}\right)^4.
\ee
Although we will focus on the scalar case below, for completeness we note that if instead $\ch_{(1,2)}$ are Majorana fermions, with
\be
 {\cal L}_{\rm int}^{\rm fermion} = - \lambda h \ch^{0 T}_1 C\ch^0_2,
 \ee
 the decay rate in the same limit takes the form,
 \be
  \Ga^{\rm fermion}_{e^+e^-} = \frac{m_e^2 \lambda^2}{480\pi^3 v^2} \left( \frac{\De m^5}{m^4_h}\right).
  \ee
One should keep in mind, of course, that in the case of fermionic singlets 
a new mediator scalar field is required alongside the Higgs in 
order to maintain SM gauge invariance \cite{chris2}.   
 
  At this point it is worth commenting on a couple of important aspects of  Higgs-mediation for a small MeV-scale energy release.
  Note first of all that the scaling of the 
  decay width changes from 
$(\Delta m)^5 v^{-4}$, as is typical for weak decays, to $(\Delta m)^5 m_e^2v^{-6}$ due to the additional 
  small factor of $m_e/v$ in the electron-Higgs vertex. Second, and an important feature in the present context, 
   is that the corresponding direct photon production through the 
  induced $hF_{\mu\nu}F^{\mu\nu}$ vertex is suppressed relative to $e^+e^-$ by the corresponding loop factor. This statement
  relies on the fact that for the ${\cal O}({\rm MeV})$-scale momentum transfer in this corner of phase space there is no Yukawa
  suppression. Indeed, given that 
  \be
   \frac{\Ga_{\gamma}}{\Ga_{e^+e^-}} \sim \left(0.01\times \frac{\om} {m_e}\right)^2,
  \ee
  with $\om$ the characteristic energy of the $\gamma$-quanta in the decay and $10^{-2}$ denoting the loop suppression factor, 
  we can estimate that the momentum scale above which the photon decay rate will exceed that of
  $e^+e^-$ is roughly 50~MeV. However, since the constraints on the subsequent production of $\gamma$'s by the
  positrons already imply a bound of ${\cal O}(10)$~MeV, this means that the $\gamma$-background is well under control
  in the interesting region of parameter space.
  
  Inserting the decay rate into (\ref{flux}), for a 100\,GeV scalar WIMP, 
   we conclude that for the Higgs-mediated decay channel to produce the required 511~keV flux, we require
   \be
    \lambda \sim 10^{-6}\;\;\;\;\;\; \Rightarrow \;\;\;\;\;\;\; (\lambda v) \sim m_e, 
   \ee
   so that the characteristic mass scale in the $h\ch_2^0\ch_1^0$ vertex is on the order of the electron mass. 
   
   To construct a simple UV-complete 
model satisfying these requirements, consider the following interaction Lagrangian,
   \be
    {\cal L}_{\rm int} =  -\left(\frac{1}{2}m^2_{ij} +\om_{ij} H^\dagger H\right) \tilde{\ch}_i\tilde{\ch}_j - \tilde{V}(\tilde{\ch}_i),\label{model}
   \ee
   with a $\mathbb{Z}_2$ symmetry $\ch_i \rightarrow - \ch_i$, and an unspecified self-interaction potential for the 
   WIMPs $\tilde{V}(\tilde{\ch}_i)$. In mass-eigenstate basis we have,
   \be
    {\cal L}_{\rm int} = - \frac{1}{2} m_i^2 \ch_i^2 -  \lambda_{ij} v h  \ch_i \ch_j  - V(h^2,\ch_i),
   \ee
   and we require that the mass spectrum is tuned to produce a small splitting $\De m = m_2 - m_1 \sim {\cal O}({\rm MeV})$. We also require
   that the residual Higgs coupling is slightly off-diagonal, where from above we need $\lambda = \lambda_{12} + \lambda_{21} \sim 10^{-6}$. The diagonal
   elements are fixed by requiring the correct cosmological abundance. This was determined in \cite{bpv} for a single scalar, and generalizing
   to the present case, for $m_i \sim 100\,{\rm GeV}$, we find that the Higgs coupling is determined as follows, 
   \be
    \lambda_{ij} \sim  \left(\begin{array}{cc} 10^{-2} & 10^{-6} \\ 10^{-6} & 10^{-2} \end{array}\right),
   \ee
   so that the required tuning in the couplings $\lambda_{12}/\lambda_{ii} \sim 10^{-4}$ is similar to that in the
   mass-splitting $\De m/m_i \sim 10^{-5}$.
      
   This leads to a rather simple model that can simultaneously  provide a WIMP state with the requisite cosmological abundance, with
   annihilation through the diagonal Higgs coupling, while having a long-lived nearby metastable state whose decay to the ground state
   can produce sufficient positrons to explain the 511~keV signal.  In a sense, it is remarkable that the sensitivity to the 511~keV flux is 
   sufficient to probe certain Higgs couplings of the WIMP down to the level of $10^{-6}$. Of course, successfully 
   fitting the overall strength of the signal is only a partial solution to the problem of the 511~keV line. Another important issue relates to the 
   spatial (i.e. radial) distribution of the source, and indeed a strong cusp,  in excess of standard NFW scaling, 
   in the dark matter distribution at $r\to0$ is usually required in order to make the decaying dark matter hypothesis consistent with the signal \cite{pp,BJ}. 
   In this regard, we should also point out that searches \cite{SPI2} for an analogous line from  the Sagittarius dwarf galaxy have not observed a signal with
   the strength one would anticipate \cite{hooper04} from a such a cuspy dark matter profile.

   \section{Concluding Remarks}
   
   We have analyzed a number of possible scenarios within which generic WIMP dark matter could provide the primary source of the galactic
   511~keV INTEGRAL/SPI signal, assuming that the associated positron production arises through MeV-scale degeneracies in the dark sector.
   Our conclusions for collisional excitation of WIMPs to a nearby excited state, and radiative capture by nuclei in the galactic interstellar medium
   were essentially negative, albeit for different reasons. Collisional excitation requires a large cross-section to produce the necessary flux, one that is
   well above the unitarity bound, and thus would need to be saturated by highly peripheral large $l$ collisions which seems implausible. The 
   radiative capture, or `recombination', of the WIMP with nuclei in the galactic medium is an interesting
   possibility that appears marginally viable in terms of producing the required photon flux, but is seemingly ruled out by a number of
   terrestrial searches, e.g. for heavy isotopes. 
   
   We are left with the third option, namely the decay of  a metastable state, with a lifetime in the range of  $10^9$ -- $10^{13}$
  years, which as argued in the preceding section appears viable, but requires a certain level of tuning in the interaction of the WIMP states
  with the Standard Model, which goes beyond the apparent tuning in the small mass splitting. It would clearly be 
  interesting to explore a more natural model of the kind outlined in Section~4, within
  which the required couplings to the Higgs may be suitably protected  from radiative destabilization, e.g. by quartic interactions
  in the potential, and specifically models of this type with additional experimental signatures. Its interesting that 
   if we were to consider a model with WIMP states $\ch_i$ which carry some Standard Model,
  i.e. weak, charge then it appears very difficult in practice to reconcile the dual constraints of an ${\cal O}({\rm MeV})$-scale momentum 
  transfer with the required lifetime.   It would therefore be of interest to see if such a scenario is necessarily limited to SM singlets, or 
  allows charged states only through the inclusion of additional mediators to the SM sector. 
      
  \subsection*{Acknowledgements}
  We would like to thank S. Ellison, J. Navarro, D. Vandenberg, M. Voloshin, and N. Weiner for helpful discussions and/or correspondence. The work 
  of MP and AR was 
  supported in part by NSERC, Canada. 
  MP acknowledges the hospitality of the Aspen Center for Physics during the summer of 2006 while this work was in gestation. Research 
  at the Perimeter Institute is supported in part by the Government of Canada through NSERC and by the Province of Ontario through MEDT.

\end{document}